\documentclass[aip,rsi,reprint,graphicx]{revtex4-1} 

\usepackage{graphicx}
\usepackage{color}
\usepackage{mathtools}
\usepackage{amssymb}

\begin{document}

\title{Highly stable polarization independent Mach-Zehnder interferometer}

\author{Michal Mi\v{c}uda}
\email[]{micuda@optics.upol.cz}

\author{Ester Dol\'akov\'a}

\author{Ivo Straka}

\author{Martina Mikov\'a}

\author{Miloslav Du\v{s}ek}

\author{Jarom\'{\i}r Fiur\'a\v{s}ek}

\author{Miroslav Je\v{z}ek}
\email[]{jezek@optics.upol.cz}

\affiliation{Department of Optics, Faculty of Science, Palack\'y University, 17. listopadu 1192/12, 77146 Olomouc, Czech Republic}

\begin{abstract}
We experimentally demonstrate optical Mach-Zehnder interferometer utilizing displaced Sagnac configuration to enhance its phase stability. The interferometer with footprint of 27$\times$40~cm offers individually accessible paths and shows phase deviation less than 0.4 deg during a 250~s long measurement. The phase drift, evaluated by means of Allan deviation, stays below 3 deg or 7 nm for 1.5 hours without any active stabilization. The polarization insensitive design is verified by measuring interference visibility as a function of input polarization. For both interferometer's output ports and all tested polarization states the visibility stays above 93\%. The discrepancy in visibility for horizontal and vertical polarization about 3.5\% is caused mainly by undesired polarization dependence of splitting ratio of the beam splitter used. The presented interferometer device is suitable for quantum-information and other sensitive applications where active stabilization is complicated and common-mode interferometer is not an option as both the interferometer arms have to be accessible individually.
\end{abstract}

\maketitle

Mach-Zehnder interferometer \cite{MachZehnder} is an essential tool
for many applications as well as in fundamental research \cite{Suda2013}.
For instance, it can be used for an indirect measurement of any physical quantity
that can be coupled to relative phase of the two interfering optical signals.
This often requires the interferometer's arms to be individually accessible, while
the interference contrast and phase stability represent further necessities.
For telecom devices \cite{modulators1,modulators2} and sensors
\cite{Washburn2011} we often prefer integrated optical circuit
implementation where interference contrast and phase stability are guaranteed
by mode coupling between waveguides and inherently monolithic design, respectively.
Free-space bulk-element interferometers, on the other hand, are highly configurable
devices but far less stable compared to the integrated ones.

Various techniques have been adopted to stabilize the relative phase
in Mach-Zehnder interferometric scheme. Thorough isolation against
environmental noise could be quite efficient in lowering the phase
uncertainty but requires a device optically contacted on ultra-low expansion
material held in vacuum \cite{Yoshito2009}. Though giving unbeatable
phase stability, such a design limits working space and flexibility of the
interferometer. Various methods of an active phase lock can be used
instead to keep the phase locked to a particular setpoint employing
a feedback loop, which has to be faster than the typical phase drift.
Unfortunately, the overall noise of the feedback loop used ultimately
limits the minimum attainable phase uncertainty to a few degrees
\cite{Furusawa2006,Furusawa2007}.
For low-level light applications, which typically employ single
photon detectors with discrete output, the response of the feedback
loop is superimposed with Poissonian photodetection noise
\cite{Huntington2005}.
To solve this issue, a strong probe signal can be used to sample
the phase drift and recover the setpoint. During this stabilization
stage, the single photon detectors have to be gated off or otherwise
isolated from the probe signal \cite{Weid2011a,Xavier2013}.
Alternatively, a faint probe signal can be used, the optical power of which
is acceptable for single photon detectors, at the expense of
increasing the stabilization stage duration. Such single-photon 
level phase-stabilization loop is typically slower than 0.1~s
\cite{Huntington2005,Makarov2004,Bartuskova2006,Bartuskova2007,Mikova2012}.

Instead of resorting to an active approach, we can exploit intrinsic
stability of some interferometric configurations. Sagnac interferometer
\cite{Sagnac1913}, for example, is well known for its inherently
stable operation. This common-mode interferometer uses the same optical
path for both interfering fields and hence the interferometer phase is
automatically stabilized. Unfortunately, there is a big disadvantage---one
cannot address individual interferometer's arm separately. An elegant
solution is to displace the arms to obtain Mach-Zehnder interferometer
where both arms can be accessed individually while maintaining Sagnac
interferometer's phase stability to a great extent \cite{Takeuchi2007}.

In many interferometric applications the polarization state of the optical
signal should be under control and the interferometer properties have
to be independent of it. Important examples include optical quantum
information protocols exploiting path as well as polarization degree
of freedom of single photons
\cite{Almeida2007,Takeuchi2009,Guo2011,Takeuchi2011,OBrien2011}
and Faraday interaction characterization \cite{LaForge2007}.
Therefore, the interferometer's action should
be verified to be polarization independent for such applications.

In this paper we describe the implementation of Mach-Zehnder interferometer
using displaced Sagnac configuration (MZDS) and verify its basic properties,
particularly interference visibility, polarization sensitivity,
and phase stability.

\begin{figure}[!bh!]
\centerline{\includegraphics[bb = 0 0 428 357, clip = true, scale = 0.54]{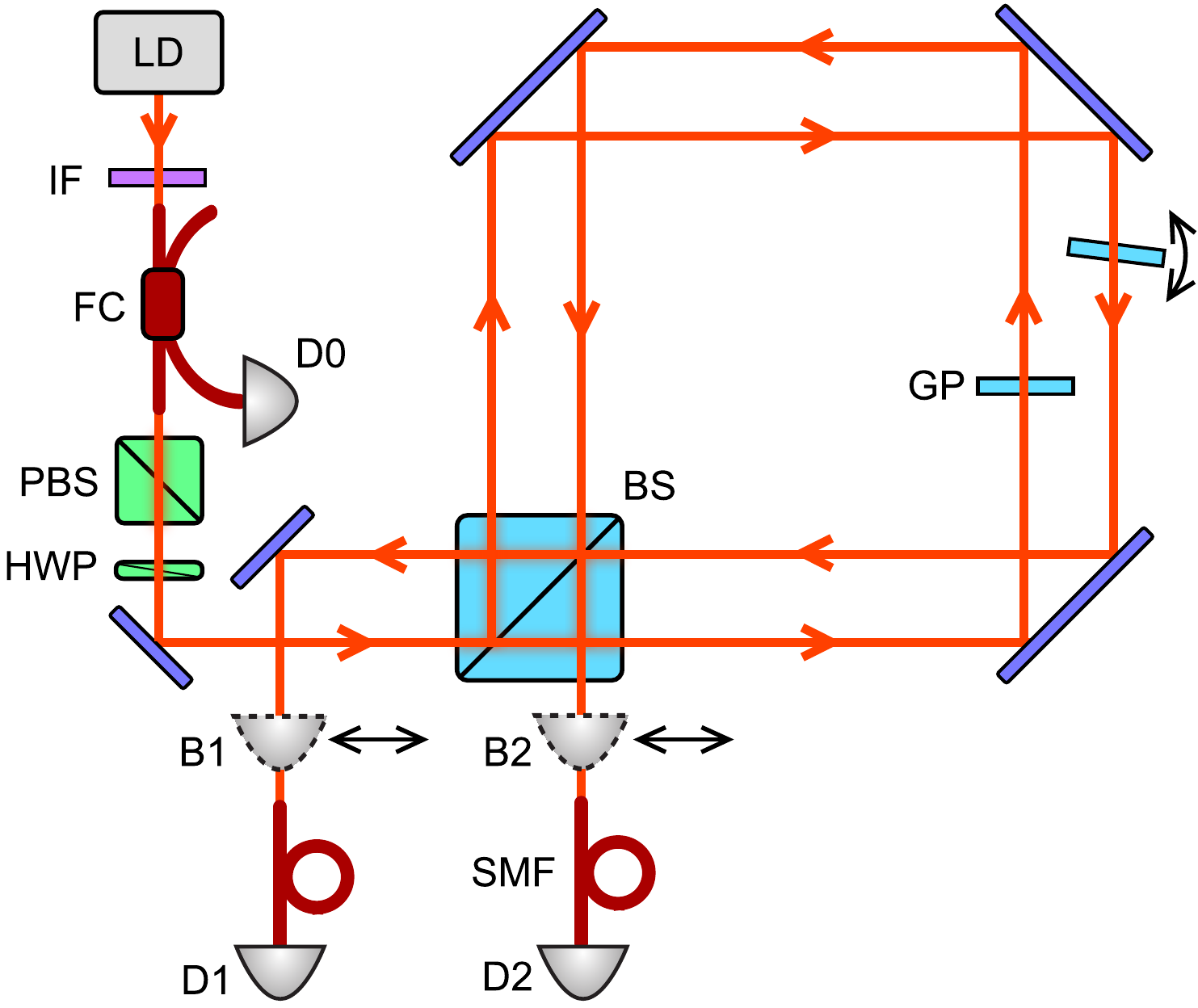}}
\caption{(Color online) Experimental realization of Mach-Zehnder interferometer
using displaced Sagnac geometry. Polarized laser light is injected
to the setup, split and superimposed again by the single beam splitter
cube (BS), and detected at the output either by bulk photodiodes B1
and B2 or by single-mode-fiber coupled photodiodes D1 and D2
to provide a definition of the spatial mode. See text for more details.}
\label{setup}
\end{figure}

The experimental setup is shown in the Fig.~\ref{setup}.
The preparation stage consists of filtered laser source with
reference detector and linear polarization preparation.
Laser diode (OZ optics, FOSS-01-3S-5/125-810-S-1) with central
wavelength of 816~nm and full spectral width at half maximum (FWHM)
of 4~nm is spectrally filtered by narrow interference filter
(Andover Corporation) centered at 814~nm with FWHM of 2~nm to match
the design wavelength of optical components employed.
Estimated coherence length after the filter reads 0.150~mm.
Optical signal is further coupled in single-mode optical fiber
(Nufern HP780) and split by a 3~dB fiber coupler (FC) from Sifam.
One part of the signal goes to a photodiode (D0), which serves
as a reference detector of the source optical power. The rest
of the laser light passes the polarizing beam splitter (PBS)
with extinction ratio larger than 1000:1 and the half wave plate
(HWP) to set a proper polarization state. Both components
are supplied by Eksma Optics.

The MZDS interferometer itself consists of a 1'' beam splitter cube
(BS) from Optida with antireflection coated sides,
three 1'' dielectric mirrors (Thorlabs BB1-E03),
and two antireflection coated 1~mm thick glass plates (GP).
The interferometer arms are $1.34$~m long and displaced by 8~mm.
The distance between beams was chosen as a reasonable trade-off
between available clear aperture of the components and convenient
individual addressing of the beams.
The manufacturer of the beam splitter cube specifies splitting ratio
of $50:50$ for both polarization modes. Our measurement shows the
splitting ratio of $45:55$ for the horizontal polarization state
(P polarization) and $43:57$ for the vertical polarization state
(S polarization). The reflectance of the mirrors employed reveals
only negligible dependence on the input polarization state.
However, small phase shift is induced between S and P polarization modes,
which does not influence the interferometer performance as both arms
feel the same phase shift. If distortion-free propagation of the
polarization state within the interferometer is required, low-dephasing
dielectric mirrors should be employed, for example OA019 from Femto-Optics.
We have measured the phase induced between S and P modes by this mirror
to be smaller than 1~deg.
Because of displaced Sagnac configuration, it is not convenient to scan
interference fringes using piezo-crystal mounted mirror. All three mirrors
are kept fixed after the initial alignment and the relative optical phase
of the interferometer is set by tilting one of the glass plates.
Mirrors and glass plates are mounted using Newport Suprema SN100C Series
mirror mounts and connected to Newport RS-4000 optical table by 1'' diameter
65~mm high brass pedestals. The beam splitter cube is epoxy glued
directly to the pedestal.

\begin{figure}[!hb!]
\centerline{\includegraphics[bb = 0 0 485 320, clip = true, scale = 0.49]{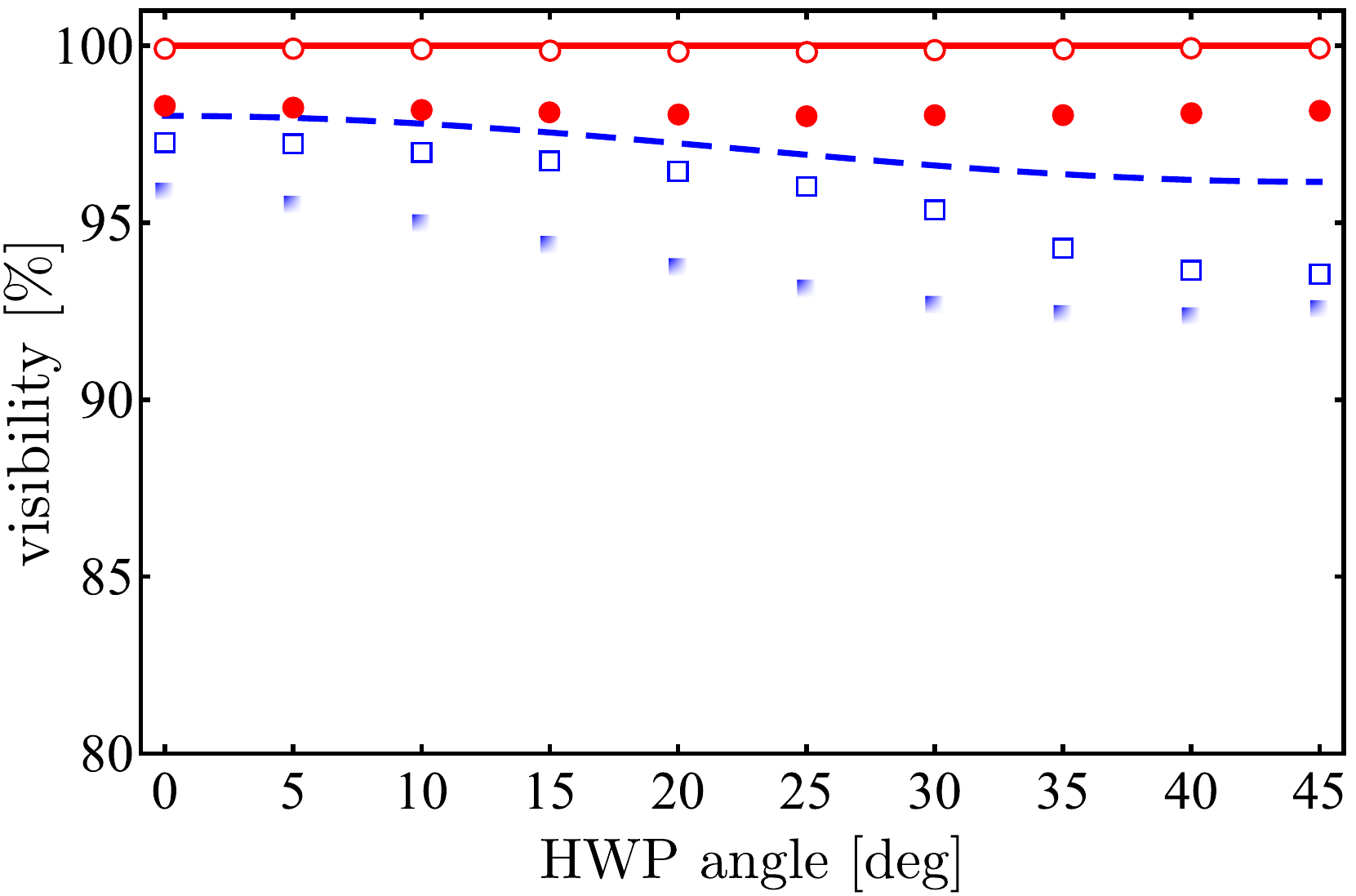}}
\caption{(Color online) The visibility dependence on the input linear polarization state:
$0\,\rm{deg}$ and $45\,\rm{deg}$ correspond to horizontal (P polarization)
and vertical polarization state (S polarization), respectively.
Curves stand for the theoretical values of visibility:
The first output port is depicted by red color (solid),
the second one by blue color (dashed).
$\color{red}{\circ}$ represents measured visibility at D1,
$\color{blue}{\square}$ represents measured visibility at D2,
$\color{red}{\bullet}$ represents measured visibility at B1, and
$\color{blue}{\blacksquare}$ represents measured visibility at B2.
Statistical errors are smaller than the symbol size.}
\label{vis_fig}
\end{figure}

The beams at interferometer's output ports are optimally coupled
into single mode optical fibers to provide a definition of the spatial
mode. These fibers are then guided to detectors. The output
optical intensity is measured by silicon p-i-n photodiodes D1 and D2
(Thorlabs DET36A). Alternatively, fiber-coupled photon counters can be
connected easily when the interferometer is operated at single photon
level. The fringe visibility \cite{PrinciplesOfOptics} is calculated
using Michelson's formula
\begin{equation}\label{VIS}
V=\frac{I_\textrm{max}-I_\textrm{min}}{I_\textrm{max}+I_\textrm{min}},
\end{equation}
where $I_\textrm{max}$ and $I_\textrm{min}$ are the maximum and
the minimum optical intensities at the particular output port,
respectively. We measured the interferometric visibility
for various angles of HWP in the
preparation stage, which corresponds to different polarization
states ranging from P to S polarizations. 
The visibility dependence on the input polarization state
is shown in the Fig.~\ref{vis_fig}.
We developed a theoretical model of the visibility
based on actual parameters of the beam splitter cube.
The measured and theoretically predicted values of the visibility
at D1 and D2 for horizontal and vertical polarization states are
summarized in the Table~\ref{vis_table1}. According to the theory
we expect the unity visibility in the first output port
for all polarization states which is in excellent agreement
with measured data---the visibility higher than 99.8\%
was reached in this port for all tested polarization states.
In the second output port, we observed approximately
0.7\% difference between theory and experiment for horizontal
polarization state and 2.5\% for vertical polarization state.
The H/V discrepancy is probably caused by a variation of the splitting
ratio of the beam splitter cube and the reflectance of the mirrors
over the active area of these components.
\begin{table}[!bh!]
\centering
\begin{tabular}{|c|c|c|c|c|}
  \hline
~ & \multicolumn{2}{|c|}{theoretical visibility [$\%$]} & \multicolumn{2}{|c|}{measured visibility [$\%$]} \\ \hline
\multicolumn{1}{|c|}{~port~~} & \multicolumn{1}{|c|}{~~~H~~~~} & \multicolumn{1}{|c|}{V} & \multicolumn{1}{|c|}{H} & \multicolumn{1}{|c|}{V} \\ \hline
D1 & ~~100~~ & 100 & $99.96\pm 0.01$ & $99.98\pm 0.01$ \\
D2 & ~~~98.02~~~ & 96.16 & ~$97.33\pm 0.03$~ & ~$93.60\pm 0.09$~\\
\hline
\end{tabular}
\caption{The theoretically estimated and the measured values of the interferometric
visibility at D1 and D2 outputs for horizontally and vertically polarized light.}
\label{vis_table1}
\end{table}
\begin{table}[!bh!]
\centering
\begin{tabular}{|c|c|c|c|c|}
  \hline
~ & \multicolumn{2}{|c|}{measured visibility [$\%$]} \\ \hline
\multicolumn{1}{|c|}{~port~~} & \multicolumn{1}{|c|}{H} & \multicolumn{1}{|c|}{V} \\ \hline
B1 & $98.35\pm 0.03$ & $98.2\pm 0.1$\\  
B2 & ~$95.93\pm 0.05$~ & ~$92.61\pm 0.04$~\\
\hline
\end{tabular}
\caption{The experimentally observed values of the interferometric visibility
at B1 and B2 outputs for horizontally and vertically polarized light,
measured by bulk photodetectors.}
\label{vis_table2}
\end{table}

For the sake of comparison, the output intensity was acquired
also by bulk photodiodes placed directly after the MZDS interferometer.
The measured values of the visibility at output ports B1 and B2
are summarized in the Table~\ref{vis_table2}. It shows that
single mode selection increases spatial mode matching
and thus the visibility for all polarization states in both output
ports by approximately 2\%.

The visibility measurement repeated on daily basis shows only negligible
variation proving its long-term stability despite of ambient temperature
fluctuations.
Interferometer phase stability is another crucial parameter which is
affected by air fluctuations, mechanical vibrations and temperature changes.
To demonstrate the phase stability of the MZDS interferometer
we acquired the intensity at detector D1 every second for ten hours.
The input polarization state was set to the horizontal linear polarization
and the initial phase was set to 0~deg, thus having interference minimum
at the output port D1. The intensity at the reference detector D0 has been
measured simultaneously. Both the photocurrents have been recorded by
12~bit data acquisition system (Pico Technology PicoLog 1216) for 10~hours. During the measurement period the temperature has been stable within 1~K.
The interferometric phase is calculated from the measured intensity at D1
corrected for the power fluctuation of the laser source.
We employ the Allan variance \cite{AllanVar} to evaluate the optimum duration
of the measurement---the integration time of the single acquisition that
yields the lowest possible phase uncertainty, which corresponds to the
minimum Allan deviation. Further, the maximum duration of the measurement
is estimated to keep the phase uncertainty reasonably small.
The Allan deviation corresponding to the integration time $\tau$ reads
\begin{equation} \label{Allan}
  \sigma(\tau)=\sqrt{\frac{1}{2N}\sum_{n=1}^{N-1}
  {\left(\bar{y}_{n+1} - \bar{y}_{n}\right)^2}},
\end{equation}
where the overall time $T$ of the long-term phase stability measurement
is divided into $N$ intervals, $T = N \tau$, and the average phase
value $\bar{y}_n$ is computed in each interval.

\begin{figure}[!th!]
\centerline{\includegraphics[bb = 0 0 490 315, clip = true, scale = 0.49]{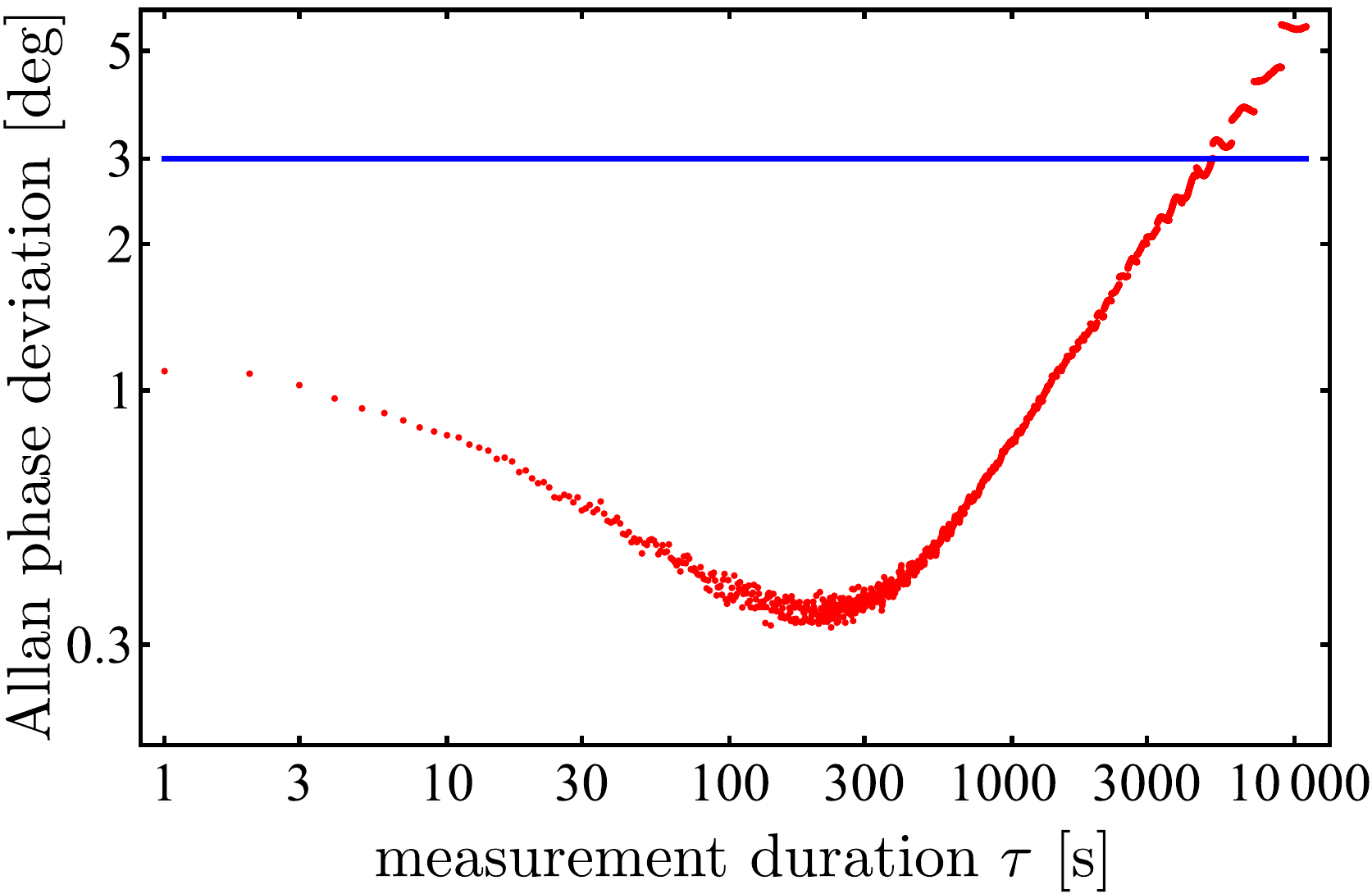}}
\caption{(Color online) Phase stability of the Mach-Zehnder interferometer using displaced
Sagnac configuration. Red dots denote measured Allan phase deviation
as a function of integration time. Blue solid line represents 3~deg
deviation level. See the text for details.}
\label{allan_fig}
\end{figure}

Typical results of the phase stability of the MZDS interferometer
are shown in Fig.~\ref{allan_fig}. The minimum Allan deviation less
than 0.39~deg is demonstrated for the integration time of about 250~s
with no stabilization technique employed. This phase uncertainty
corresponds to the interferometer arm length deviation of 0.87~nm
and the relative length deviation of $0.65\times10^{-9}$.
If we allow ourselves to have the phase deviation less than 3~deg
then the duration of the measurement can be extended up to 1.5 hours.
The measured phase uncertainty of the MZDS interferometer can be
compared with phase stability of Mach-Zehnder interferometer
for heterodyne metrology presented by Niwa and collaborators
\cite{Yoshito2009}.
They were able to keep the interferometer arm length deviation below
20~pm over a hour. This excellent stability figure was achieved
by building their interferometer on ultra-low expansion glass base
plate with dimensions of $5\times5$~cm and placing the whole
setup into a vacuum chamber with temperature stabilization within
1~mK over four hours. The resulting relative length deviation
of about $0.2\cdot 10^{-9}$ per hour is better by a factor of ten
than what we have shown here for the MZDS interferometer
operating under much less demanding conditions.
Further, the observed passive phase stability is comparable
with an active phase lock loop, the noise of which can easily
induce the overall phase uncertainty of an order of 1~deg
even if measured at high frequency where the ambient noise
is significantly lower. Takeno et al \cite{Furusawa2007}
showed the actively locked phase uncertainty of 1.5~deg
at 1~MHz sideband with the bandwidth of 30~kHz. Eberle et al. \cite{Schnabel2013} demonstrated a sub-degree phase uncertainty at 8 MHz sideband with the bandwidth of 200 kHz for up to 15 minutes. Phase noise suppression at acoustic and lower frequencies
and for longer times is typically not crucial for quantum
continuous-variables experiments with sideband encoded information
but it is generally very important for single photon level
interferometric measurements with prolonged data acquisition
times \cite{Takeuchi2007}.

In conclusion, we have reported the inherently stable Mach-Zehnder interferometer
in displaced Sagnac geometry, examined its visibility, polarization sensitivity
and phase stability.
The visibility of the interferometer slightly depends on the polarization state.
We have measured approximately 3.5\% difference in visibility between the S and P
polarization modes at one output port of the interferometer and virtually no
difference at the other one. This behavior agrees reasonably with the simple
theoretical model developed, which takes into account measured splitting ratios
of the beam splitter cube for both the polarization modes.
With more balanced beam splitter the overall performance of the interferometer
will improve as can be already seen at its first port where the visibility
does not depend on the splitting ratio and is given solely by mode matching.
At this port we consistently observe the interference visibility higher
than 99.8\% with negligible dependence on the input polarization.
Further, we have explored long-term stability of the interferometric phase.
We have demonstrated the root-mean-square phase noise of the MZDS interferometer
as low as 0.39~deg within the measurement time of 250~s and of about 3~deg
when integrated from 0.2~mHz to 1~Hz. We expect no significant phase noise for
higher frequencies because of the closeness of MZDS interferometer's paths
and thus its virtual immunity to acoustic waves with frequencies up to tens~kHz.
The inherent robustness and notable long-term passive phase stability make
the MZDS interferometer good candidate to encode a spatial quantum bit
carried by a single photon in various quantum information protocols.
Together with low polarization sensitivity and ease to address
its individual arms, it enables the hyper-encoding using both
the spatial and polarization degrees of freedom.
Alternatively, Jamin--Lebedeff interferometer \cite{JaminLebedeff}
formed by a pair of calcite prisms can be used for applications
where the polarization insensitivity is not required \cite{White2003}.
The presented interferometer device thus seems suitable for
quantum-information and other sensitive applications where active phase
stabilization is complicated and common-mode interferometer is not
an option as both interfering arms have to be accessible individually.

\acknowledgments
This work was supported
by the Czech Science Foundation (GACR 13-20319S)
and by the Palack\'y University (IGA-PrF-2014008).

\end{document}